\documentclass[
aps,%
12pt,%
final,%
notitlepage,%
oneside,%
onecolumn,%
nobibnotes,%
nofootinbib,%
superscriptaddress,%
noshowpacs]%
{revtex4}
\usepackage{graphicx}
\usepackage{color}
\usepackage{xspace}

\newcommand{\IXPE}{\textit{IXPE}\xspace}

\newcommand{\Nustar}{\textit{NuSTAR}}
\newcommand{\NICER}{{NICER}}

\begin{document}

\noindent {\it Astronomy Reports, 2023, Vol. , No. }
\bigskip\bigskip  \hrule\smallskip\hrule
\vspace{35mm}


\title{Polarised light from accreting low mass X-ray binaries \footnote{Paper presented at the Fifth Zeldovich meeting, an international conference in honor of Ya. B. Zeldovich held in Yerevan, Armenia on June 12--16, 2023. Published by the recommendation of the special editors: R. Ruffini, N. Sahakyan and G. V. Vereshchagin.}}

\author{\bf \copyright $\:$  2023.
\quad \firstname{F.}~\surname{Capitanio,}}%
\email{fiamma.capitanio@inaf.it}
\affiliation{INAF Istituto di Astrofisica e Planetologia Spaziali, Rome, Italy}%

\author{\bf \firstname{A.}~\surname{Gnarini}}
\affiliation{Universit\'a degli Studi Roma Tre, Rome, Italy}

\author{\bf \firstname{S.}~\surname{Fabiani}}
\affiliation{INAF Istituto di Astrofisica e Planetologia Spaziali,  Rome, Italy}
\author{\bf \firstname{F.}~\surname{Ursini}}
\affiliation{Universit\'a degli Studi Roma Tre, Rome, Italy.}
\author{\bf \firstname{R.}~\surname{Farinelli}}
\affiliation{INAF Osservatorio di Astrofisica e Scienza dello Spazio di Bologna, Bologna, Italy}
\author{\bf \firstname{M.}~\surname{Cocchi}}
\affiliation{INAF Osservatorio Astronomico di Cagliari, Cagliari, Italy}
\author{\bf \firstname{N.}~\surname{Rodriguez Cavero}}
\affiliation{Washington University in St. Louis, United States}

\author{\bf \firstname{L.}~\surname{Marra}}
\affiliation{Universit\'a degli Studi Roma Tre, Rome, Italy}
\begin{abstract}

\centerline{\footnotesize Received: ;$\;$
Revised: ;$\;$ Accepted: .}\bigskip\bigskip\bigskip

Thanks to \IXPE, the X-ray spectro-polarimeter launched at the end of 2021, X-ray polarimetry has finally become an extraordinary tool in investigating the physics of accretion in low mass X-ray binaries. Similarly to what happened with gravitational waves, X-ray polarimetry would play a
new complementary but at the same time fundamental role in the high-energy astrophysical domain. We summarize here the first 1.5 year results on accreting low-mass X-ray binaries obtained by a huge \IXPE\ observation campaign coordinated with the principal X-ray and $\gamma$-ray telescopes. Then we compare these results with the theoretical prediction highlighting the unexpected results

\end{abstract}

\maketitle

\section{Introduction}

Accreting Low-mass X-ray binaries (LMXBs) host a compact object (non-pulsating Neutron Star, NS or a Black Hole, BH) accreting via Roche-lobe overflow from a low mass star companion. They are highly variable objects on time scales from months down to fractions of a seconds. They are classified depending on the tracks they draw on the color-color diagram (CCD) or  hardness-intensity diagram (HID) if they are NS and BH, respectively. The majority of BH-LMXB are transient sources, while NS-LMXB can be either persistent or transient.

Their spectral and timing behaviors during the outburst generally follow the same pattern of transition between hard and soft states, even though exceptions have been occasionally reported in literature \cite{1,2}. The X-ray spectral-timing properties of LMXBs show two alternating main states: Low-Hard (LHS) and High-Soft (HSS) \cite{3}, passing through two different intermediate states\footnote{intermediate states are commonly found only in BH LMXB} having similar X-ray spectra but very different radio and timing properties. Both main states of NS and BH LMXB exhibit a a spectral continuum usually described by a combination of a soft thermal component plus a harder electron scattering one with reflection by a cold medium. The HSS presents instead peculiar differences between NS (still described by Comptonization of a relatively cool thermal electron population ($kT \sim$3 keV) at high optical depth) and BH (with a power law-like component extending to hundreds of keV appearing to be consistent with a hybrid or non-thermal electron distribution).

The X-ray spectroscopy of LMXB provides information on the physical parameters of the emitting regions, while polarimetry, not available before \IXPE, about geometrical properties such as the shape and extension of the region where Comptonization occurs. In fact, different geometries result in quite different polarization degrees \cite{4}. Moreover, the parallel transport in the warped space-time close to the central object is expected to rotate the polarization vector, leading to a swing in polarization angle across the disc energy spectrum in the \IXPE band \cite{5}. Since the amplitude of this effect depends on BH spin, X-ray polarimetry of the BH-HSS should in principle prodide a novel spin diagnostic. This new method is independent with respect to the other existing methods often providing inconsistent results (see \cite{6} and references therein).

In this context, it is fundamental to properly determine the spectral model to disentangle different components in order to derive their polarization in \IXPE\ data from 2 to 8 keV. \IXPE\ alone is not sufficient to do this, because of the limited energy band-pass (2-8 keV) and energy resolution. Broad-band spectroscopy is mandatory through coordinated observations with other X-ray facilities.
We summarize in the following sections some of the spectro-polarimetric results obtained by a huge campaign of simultaneous observations between \IXPE\ and the principal X-ray and $\gamma$ telescopes for Z, Atoll and black hole low mass X-ray binaries. 

\subsection{IXPE}
The \textit{Imaging X-ray Polarimetry Explorer}, \IXPE\ \cite{7} is a NASA/ASI mission launched on 2021 December 9 providing space, energy and time resolved polarimetry \cite{8}.
With respect to the previous X-ray polarimetric mission, {OSO}-8, \IXPE\ needs about two orders of magnitude less exposure time to reach the same sensitivity, and it provides imaging capability with $\leq30$  arc-second angular resolution over $>11$ arc-minutes field of view, together with 1--2\,$\mu$s timing accuracy and a moderate spectral resolution typical for proportional counters.
It consists of three X-ray telescopes with identical mirror modules and identical polarization-sensitive imaging detector units (DUs) at their focus. 

\section{Results}
Neutron star low-mass X-ray binaries (NS-LMXBs) are among the X-ray astronomical objects that \IXPE\ has investigating during the first 1.5 year of observation campaign. The already observed Z-sources are:
Cyg X$-$2, GX 5$-$1  \cite{9,10}, the peculiar Z-Atoll transient XTE J1701$-$462 \cite{11} and recently Sco X$-$1 (paper in preparation). While the already observed Atoll sources are: GS 1826$-$238, GX 9+9 \cite{12,13}, the ultra compact X-ray binary 4U 1820$-$30 \cite{14} and recently 4U 1624$-$49 (paper in preparation). Concerning the BH-LMXB, the transient nature of the majority of these kind of objects has given the possibility  to observe with \IXPE\ a smaller number of sources: the transient LMXB, 4U 1630$-$47 \cite{15,16}, the persistent LMXB, 4U 1957+115 (paper in preparation) and very recently the new discovered BH-XRB Swift J1727.8$-$1613 (paper in preparation).

\subsection{Z sources}
 
\begin{figure}
\includegraphics[width=0.4\textwidth]{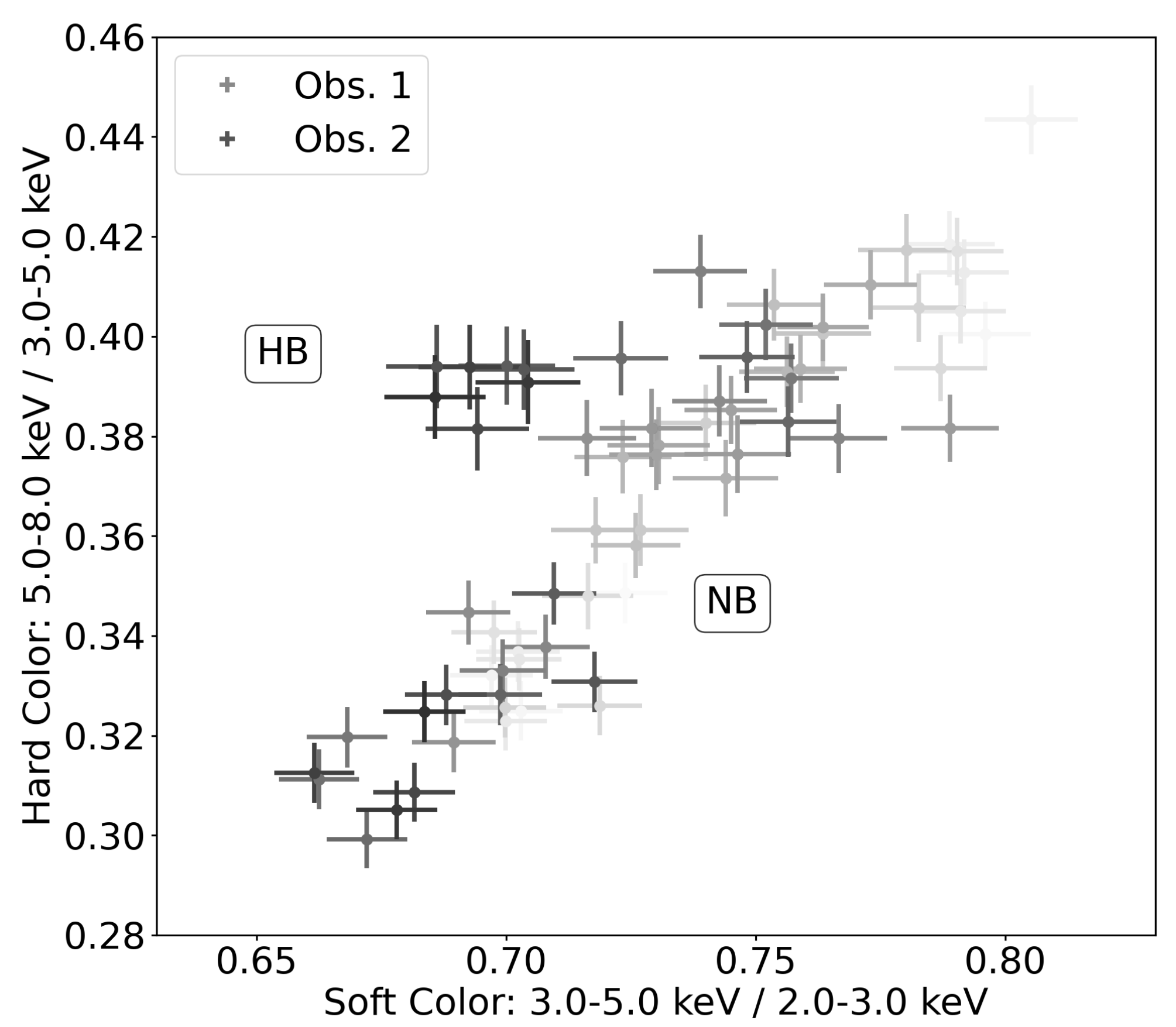}
\includegraphics[width=0.35\textwidth]{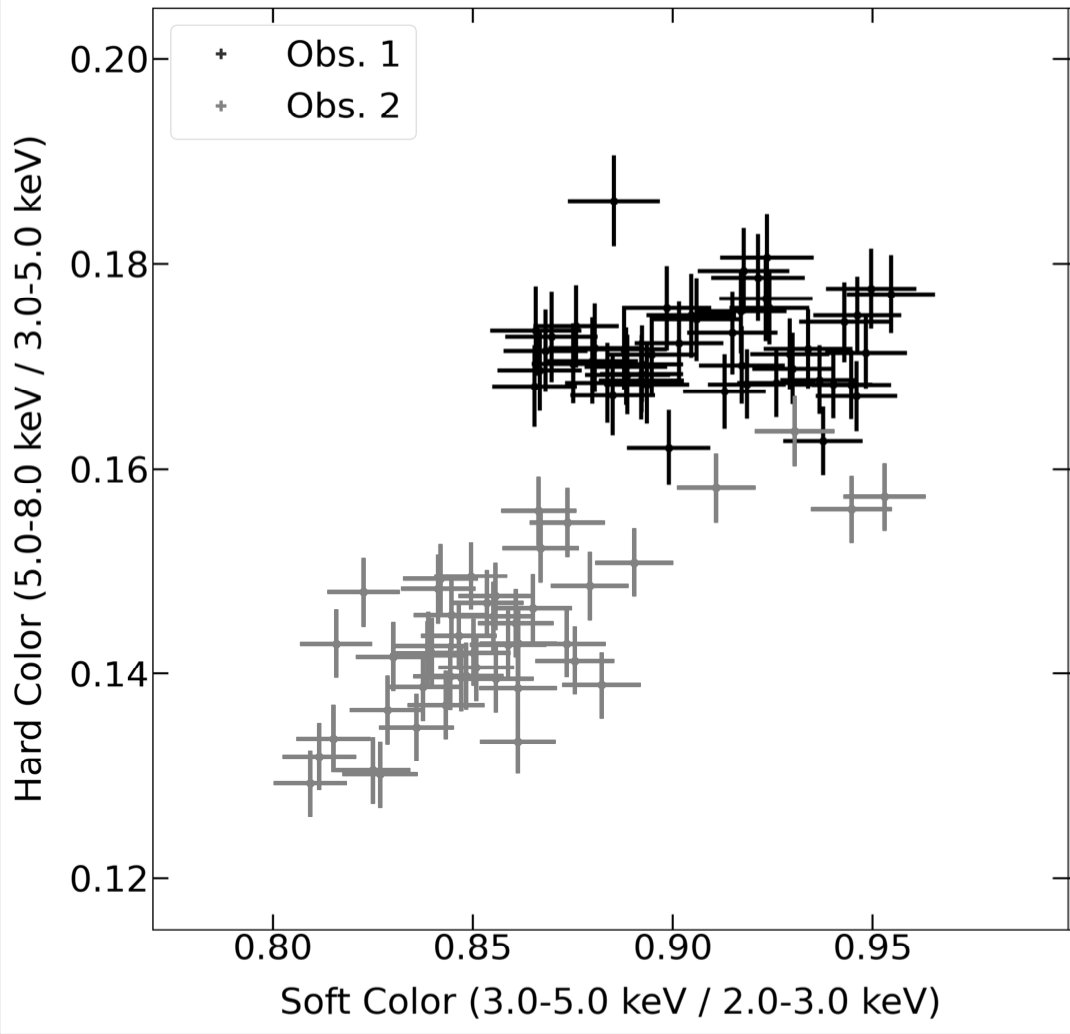}
\caption{ Left panel: \IXPE\ CCD of XTE J1701$-$462 (see ~\cite{2} during the first \IXPE\ Observation (HB) and during the second \IXPE\ observations (NB). Colors darken with increasing time in order to show the trend of the patterns. Each point corresponds to 500 s integration time. Right panel: \IXPE\ CCD of GX 5$-$1 during the first \IXPE\ Observation (black points) and during the second \IXPE\ observations (red points) Each point corresponds to 450 s integration time (for details see \cite{10}. }
\label{ccd1701}
\end{figure}

XTE J1701$-$462 was observed twice by \IXPE: on 2022 September 29 and 10 days later. During the first pointing, an average 2--8 keV polarization degree of (4.6$\pm$0.4)\% was measured. Conversely, only a $\sim$0.6\% average degree was obtained during the second pointing. As the left panel of Figure~\ref{ccd1701} clearly shows, the source, between the two different observations moved from horizontal branch (HB) to normal/flaring branch (NB/FB) of the Z-track in the CCD~\footnote{IXPE is not able to distinguish between FB and NB}. The same behaviour was observed in GX 5$-$1: the right panel of Figure~\ref{ccd1701} shows the CCD diagram of the \IXPE\ data during the two observations. The shape of the CDD was confirmed also by simultaneous \NICER\ and \Nustar\ observations for both sources as reported in \cite{17} and \cite{10}.
In particular, in GX 5$-$1, the X-ray polarization degree was found to be 3.7\% $\pm$ 0.4\% (at 90\% confidence level) when the source was in the HB 
and 1.8\% $\pm$ 0.4\% when the source was in the NB/FB. Cyg X$-$2 has been observed only in the NB with a {\it PD}=1.85\% \cite{9}. These results indicate that the variation of polarization degree presents a strong dependence from the position of the source in CCD. In Table~\ref{Z} we resume the values of {\it PD} and {\it PA} observed in Z-sources until now. For example, the {\it PD} observed in Cyg X$-$2 in HS is consistent with the one measured in the same state in GX 5$-$1, this means that the { \it PD} measured in CyG X$-$2 could be the lowest possible. The differences in {\it PD} values between HB and NF/FB in XTE 1701$-$462 are greater than those observed in GX 5$-$1.

Before the \IXPE\ launch, Long et al. 2022~\cite{19} published a significant detection of a polarization signal in Sco X$-$1 in the energy range 4--8 keV, when the source was probably in the HB/FB with the {\it PolarLight}~\citep{20} balloon experiment. 
Evidence for variation of the polarization angle $\sim$20 deg with energy has been found in GX 5$-$1, likely related to the non-orthogonal polarization angles of the disk and Comptonization components which peak at different
energies. While Cyg X$-$2 \cite{9} and Sco X$-$1 \cite{19} show a polarization angle oriented in the same direction of the radio jets. Concerning GX 5$-$1 and XTE J1701$-$462, it is not possible to verify this hypotheses because the direction of their radio jets is still unknown.

\subsection{Atoll sources}
In Atoll sources the polarization seems to be generally rather weak (less than 1.5\% in the total energy range, \cite{12,13}) but with a strong energy dependence. For example, in the case of GX9$+$9 the average in 2-8 keV {\it PD} is 1.4$\pm$0.3. Dividing the energy range in two parts (2-4 keV and 4-8 keV), the source {\it PD} is constrained only in the higher energy range, {\it PD}=2.2$\pm$0.5 \cite{13}. Moreover, in the case of ultra compact LMXB 4U 1820$-$30, the {\it PD} strongly increase in the energy range between 7-8 keV from an upper limit of {\it PD} $<$1\% to a {\it PD} that could reach even $\sim$10\%, respectively \cite{14}.
It is important to note that looking at the \NICER\ + \Nustar\ spectrum of 4U 1820$-$30 \cite{14} the {\it PD} rapidly increase as soon as the disk component drops down to a flux below 10$^{-12}$ erg cm$^{-2}$ s$^{-1}$ indicating that the {\it PA} of each single component of the spectrum (disk, Comptonization and reflection) are oriented differently resulting in a decreasing of the total {\it PD} in the spectral range where both component are present. 

\begin{figure*}
\centering
\includegraphics[width=0.5\columnwidth]{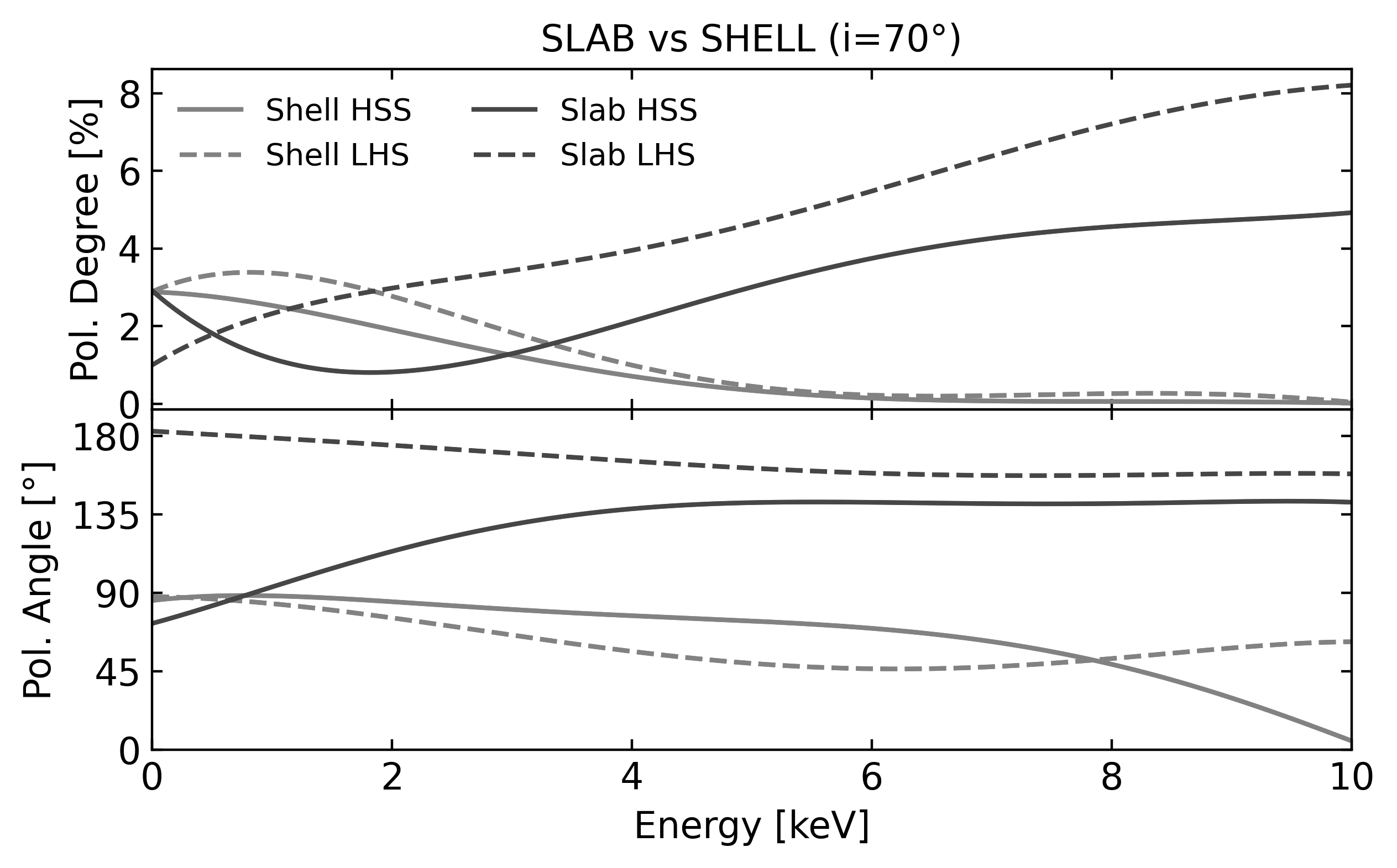}
 \caption{Polarization degree and angle assuming two different coronal geometries for a NS-LMXB in HSS or in LHS observed at 70$^\circ$ inclination \cite{21}.}
  \label{simu}
\end{figure*}

Most of theoretical works that predict the polarised fraction of light emerging from LMXBs concerns BH-LMXB. The only recent work that simulate the evolution of{\it PD} and {\it PA} in weakly magnetised LMXB has been published by Gnarini et al. 2022~\cite{21}. 
The simulations have been performed with the general relativistic Monte Carlo code, MONK \cite{22}, suitably adapted to compute the X-ray polarized radiation coming from weakly magnetized NS-LMXBs in Kerr space-time, accounting for the contributions of the disk, corona and the NS. 
As theoretically predicted, {\it PD} and {\it PA} are expected to be different from different geometries of the X-ray corona and for different viewing angles of the source. Fig.~\ref{simu} shows the variation of {\it PD} and {\it PA} for two different geometry of the Compton corona, calculated with a Monte Carlo code and including GR effects (in the Kerr metric, assuming a value of the spin of 0.16, as appropriate for a 3 ms NS rotation, see \cite{23}. Different geometries (and different levels of asphericity) result in quite different polarization degrees and orientations, which can therefore help distinguishing between them. 

\begin{figure*}

\includegraphics[width=0.4\textwidth]{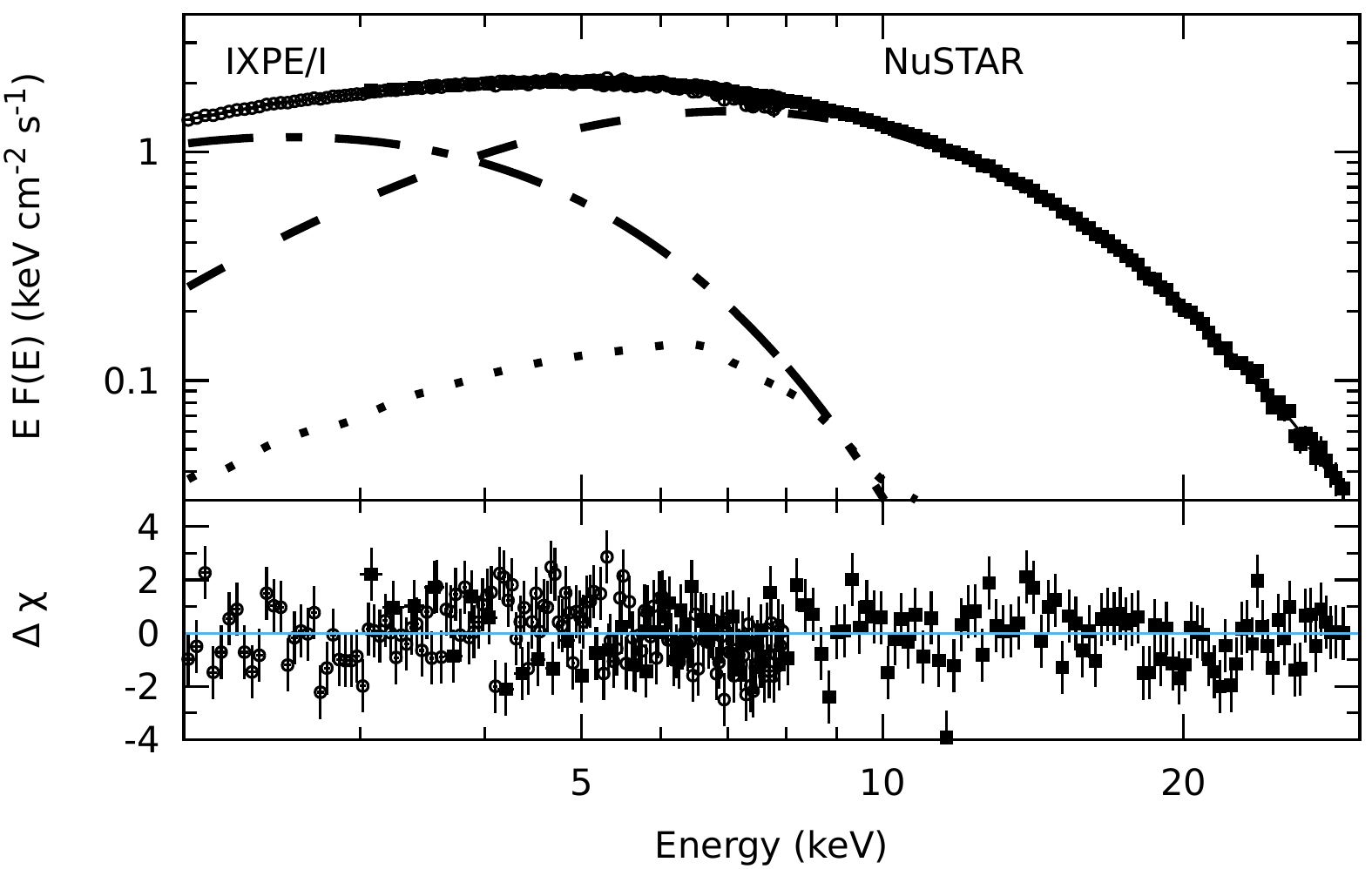}
\hspace{1cm}
\includegraphics[width=0.4\textwidth]{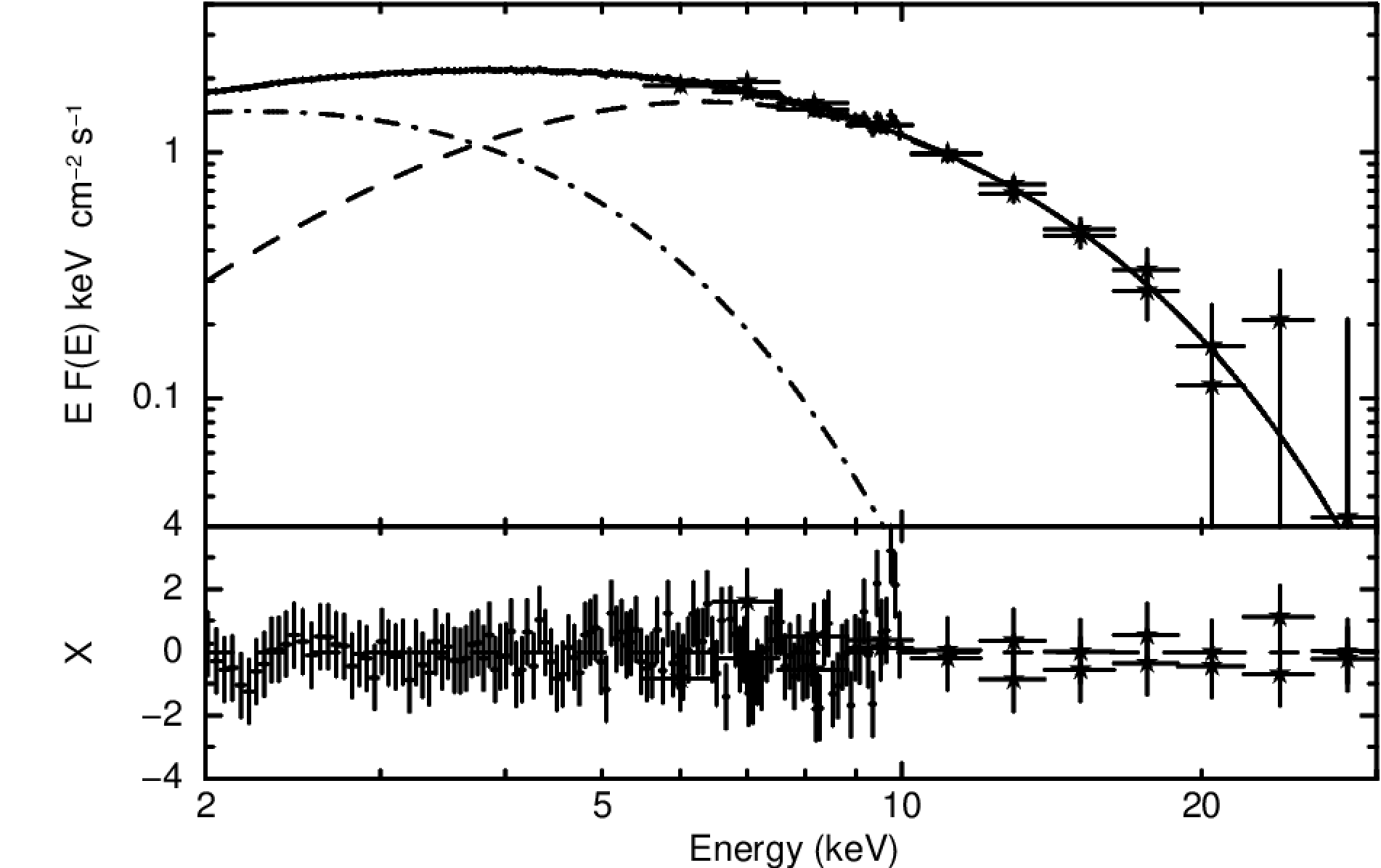}
\caption{{\bf Top panel}: deconvolved \IXPE\ and \Nustar\
spectrum of GX~9+9 fitted with a model composed of diskbb (dash-dotted line)
and comptb (dashed line)  and relxillns (dotted line). Also residuals in units of $\sigma$ are shown (for best fit model parameters see \cite{13}. 
{\bf Bottom panel}: deconvolved spectrum of GS 1826$-$238 as observed by NICER (black crosses) and JEM-X1 and 2 (blue crosses). The cyan dashed line represents the accretion disk emission (diskbb); the pink dotted–dashed line
represents the emission of the Comptonized component (comptt). Also residuals in units of $\sigma$ are shown (for best fit model parameters Capitanio et al.+2023).}
\label{gx_gs_spec}
\end{figure*}

GS 1826$-$238 and GX9+9 are the first two Atoll sources observed by \IXPE\ for 100 ks each in March and October 2022, respectively. Figure~\ref{gx_gs_spec} shows the \NICER\ + \Nustar\ (GX 9+9) and
\NICER\ + {\it INTEGRAL} (GS 1826$-$238) spectra, simultaneous with the \IXPE\ observations, of the two sources. 
The two spectra are very similar to each other: similar inclination, similar distance, similar flux and similar spectral parameters. However, GS 1826$-$238 does not show any reflection features, while for GX 9+9 a reflection component is needed in order to obtain an good fit of the data. The other difference is that that it is possible to detect a faint signal of polarization in GX 9+9 that increase with energies. While for GS 1826-238, only tiny upper limits were obtained (See Table~\ref{gx_gs_tab}). It is important to notice that {\it PD} values of GX 9+9 are close to the upper limits found in GS 1826$-$238. Therefore, if GS 1826$-$238 was polarised at the same level of GX 9+9 there should be at least a marginal detection in \IXPE. As reported by~\cite{24} and~\cite{25}, the reflection from the accretion disk  of the radiation produced by the SL or self-illumination of the disk can produce substantial polarization.
However, we do not detect in GS 1826$-$238, at least with the spectral resolution of \NICER, the iron line that is a typical signature of disk reflection in the HSS sources (e.g. Cyg X$-$2 and Sco X$-$1,\cite{26,27}. 
One possibility is that the disk is strongly ionized reducing the strength of the iron line. 
On the other hand, the latitudinal extent of the SL might not be large enough to produce significant illumination of the disk resulting in a weak signal. If we consider the latter case as the reason why the reflection is not present in GS 1826$-$238, we can deduce that the presence of detectable reflection component in GX 9+9 could be the reason why we detect an higher level of polarization in GX 9+9 respect to GS 1826$-$238.


\subsection{Black Hole Low Mass X-ray Binaries}

\begin{figure*}
\centering
\includegraphics[width=0.8\columnwidth]{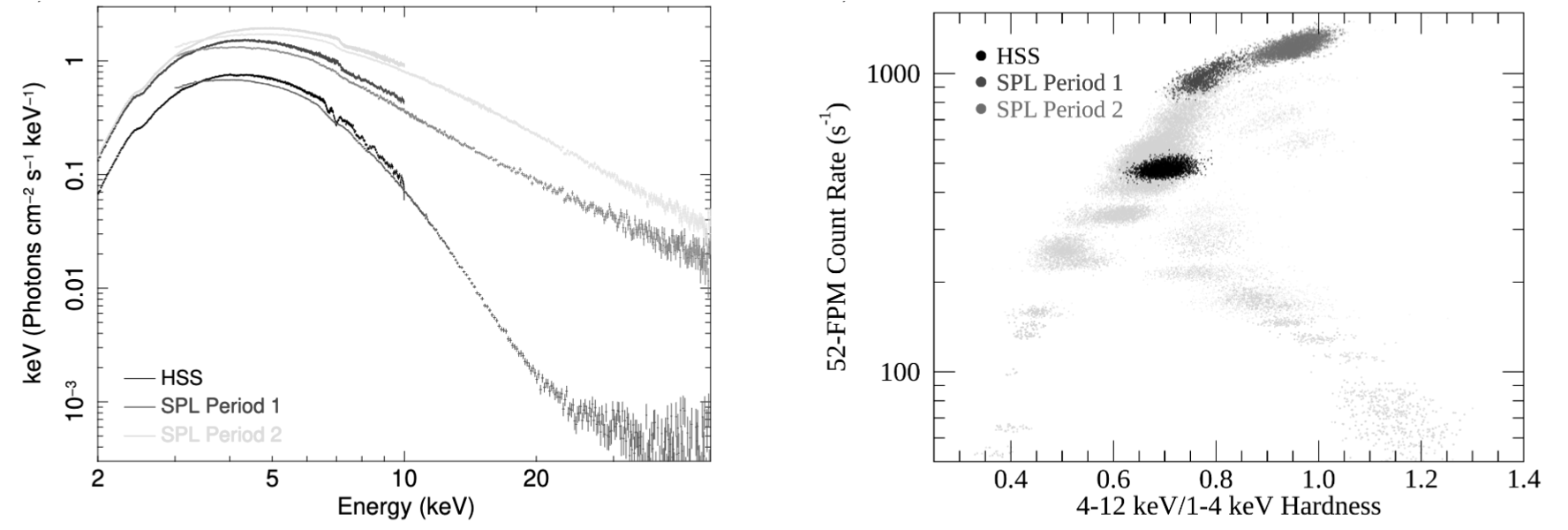}
 \caption{Figure from Rogriguez-Cavero et al. 2023 \cite{15} 
 a) NICER (2–10 keV) and NuSTAR (3–50 keV) spectra of the HSS (black) from Paper I and from the current SPL
Period 1 (gray) and Period 2 observations (light gray). The spectra were unfolded using a unit constant model for both instruments.
b) Hardness-intensity diagram from NICER data of the HSS (black) and SPL state Period 1 (gray) and Period 2 (light gray), in 8 s
intervals. Data from all previous NICER observations of 4U 1630$-$47 are shown in gray. Rates have been normalized as if all 52
of NICER’s FPMs were pointing at the source}
  \label{4u}
\end{figure*}

Connors et al 1980~\cite{28}, Dov{\v{c}}iak et al. 2008~\cite{29}, Schnittman \& Krolik 2009~\cite{25}, and Taverna et al. 2020~\cite{30}  account for scattering as the only responsible for photon polarization in accreting black holes. They predict a level of polarization of $<$5\% in the \IXPE\ range. \cite{30} included returning radiation due to strong gravity effects.
These simulations predict that the spectrum of {\it PD} and {\it PA} in function of energy depends on the spin of the BH. In particular \cite{31}, including absorption effect but without including the effect of strong gravity, show that the {\it PD} tends to be higher for high inclination sources with low spin (for which the adsorption effects are more important). For high values of the spin, the scattering effects dominate and the polarization tends to approach the Chandrasekhar profile. Moreover, the parallel transport in the warped space-time close to the central object (especially for BH) is expected to rotate the polarization vector, leading to a swing in polarization angle in function of energy. For intermediate values of the spin, this rotation is in the \IXPE\ energy band. Therefore, the X-ray polarimetry of the BH-LMXB in HSS provides a novel spin diagnostic. In LHS, on the contrary the fraction of polarised light should come from compton corona and , as in the case of NS-LMXB, should give information on the geometry of the corona itself \cite{22}.
 The 2022 outburst of the BH-LMXB 4U 1630$-$47 was observed two times: during the first observation the source was in a pure High Soft State (HSS): the spectrum was dominated by the accretion disk and the contribution from the corona emission (between 2-8 keV) is very small (less than 1\%). The second IXPE observation  wasperformed when the source showed an evident steep power law component and was divided into two distinct periods owing to the sudden increase in the source flux as showed by the  \NICER\ + \Nustar\ spectrum simultaneous with \IXPE\ in  Figure~\ref{4u}.
  4U 1630$-$47 is a peculiar source, sometimes it lacks bright hard state starting the outburst directly with a bright soft spectrum~\cite{32}. Also the 2022 outburst followed the same  peculiar behaviour as demonstrated by the \NICER\ HID in Figure~\ref{4u}. Concerning the polarised light emitted from this source, as reported in \cite{15}, the {\it PA} remains almost constant and the {\it PD} is quite high. This result can be interpret as a low spin source. An unexpected results remain the fact that the polarization remains constant as the source state varies from pure HSS to a peculiar Steep-power law state. In one of the \NICER\ observations simultaneous with \IXPE, in the typical HSS spectrum, some absorption lines that indicate the presence of wind are present.  In SPLS the spectrum changes drastically and there is not any wind detection. However, the {\it PD} and {\it PA}  remain substantially constant \cite{15}.

 Concerning the hard-states, at the date of submission of this paper, observations of a bright BH-LMXB Swift 1727.8$-$1613 are on-going. The only \IXPE\ results on a LMXB in hard  state (neither XRB hosting NS or BH) currently available are those of Cyg X$-$1~\cite{33}. Being Cyg X$-$1 a persistent high mass XRB, the accretion is also via companion star wind. The scattering from the wind of the companion star could be an additional, non-negligible, source of polarization. Instead, in the case of LMXBs in hard state (with the disk truncated far away), the principal source of polarised light (even in 2-8 keV energy range) should be due to the Comptonised radiation coming from the corona.
 The measured {\it PD} in Cyg X$-$1 is higher than pre-IXPE model predictions  and can be achieved only considering a inclination of the inner accretion disc greater than the inclination obtained from standard measurements (warped disk, \cite{33}).
 
However, the first results on Cyg X$-$1 \cite{33} seem to rule out both the jet origin of polarised light and the lamp post corona geometry and favouring an extended corona. Moreover, there is strong evidence that all LMXBs (either hosting a BH or a NS as compact object) share the same characteristic: the polarization angle is aligned to the relativistic radio jets as, for example, in the cases of Cyg X$-$1, Cyg X$-$2 and Sco X$-$1 where a past relativistic radio jet image was available in the literature highlighting the jet inclination angle on the sky ( \cite{9,19,32} and references therein). 

\section{Discussion and Conclusions}
The first 1.5 year of \IXPE\ operations disclosed a scenario more complex than expected:
For example, the predicted polarization in LMXB hosting a NS was $<$ 2\%. Instead we measured with \IXPE\ averaged {\it PD} that can reach $>$ 4\% for both Z-sources and BH XRBs. While for Atoll sources in HSS the polarization seems to be lower than predicted \cite{12,13}. The polarization is probably related to the reflection component \cite{13} and could became an important tool to investigate the presence of reflection component even in the case of the lack of any related spectral features due to totally ionized disk~\cite{10}.
Observations of Z-sources indicate that the polarization is very variable and strongly related to the position of the source in the CCD \cite{10,11}. This variation is far to be explained. However, we can suppose that it is due to a variation of the latitudinal extensions of the SL.  
GX 5$-$1 manifested an unexpected variation of the {\it PA} as a function of energy by $\sim$ 20$\deg$. This variation is likely related to the different {\it PA} of the disk and Comptonization components in this case non-orthogonal and have the emission peaks at different energies. 
Concerning the {\it PA}, until now there is an evidence that in LMXB it is aligned to the relativistic radio jets. This fact could have been verified in the cases of Cyg X$-$2, Sco X$-$1 and the BH LMXB Cyg X$-$1 because of a past relativistic radio jet image available in the literature that permitted to fix the jet inclination angle on the sky.
No hard states of NS-LMXB have been observed by \IXPE\ until the date of submission of this proceeding. In fact, the observed  Atoll sources were all in soft state because of their grater brightness respect the hard state emission. Concerning BH-LMXB the observations of a bright new transient Swift J1727.8$-$1613 are on-going.
Finally, in 4U 1630$-$476 \IXPE\ observations found an high level of polarization that is substantially independent from: spectra, wind and accretion. Probably the solution could be found in an out-flowing, partially-ionized accretion disk atmosphere  that produces the observed high {\it PD} as a result of Thomson scattering \cite{16}. Moreover it still impossible to explain why the combined thermal and power-law emission end up having similar polarization properties as the thermal emission alone (see \cite{15} for details)

\begin{acknowledgments}
The Imaging X-ray Polarimetry Explorer (IXPE) is a joint US and Italian mission.  
This research used data products provided by the \IXPE\ Team (MSFC, SSDC, INAF, and INFN) and distributed with additional software tools by the High-Energy Astrophysics Science Archive Research Center (HEASARC), at NASA Goddard Space Flight Center (GSFC).
Partly based on observations with {\it INTEGRAL}, an ESA project with instruments and science data centre funded by ESA member states with the participation of the Russian Federation and the USA. Data have been reduced using the \href{https://www.astro.unige.ch/mmoda/}{Multi-Messenger Online Data Analysis} platform provided by the ISDC (University of Geneva), EPFL, APC, and KAU. 

\end{acknowledgments}

\section*{Funding}

The US contribution is supported by the National Aeronautics and Space Administration (NASA) and led and managed by its Marshall Space Flight Center (MSFC), with industry partner Ball Aerospace (contract NNM15AA18C).
The Italian contribution is supported by the Italian Space Agency (Agenzia Spaziale Italiana, ASI) through contract ASI-OHBI-2022-13-I.0, agreements ASI-INAF-2022-19-HH.0 and ASI-INFN-2017.13-H0, and its Space Science Data Center (SSDC) with agreements ASI-INAF-2022-14-HH.0 and ASI-INFN 2021-43-HH.0, and by the Istituto Nazionale di Astrofisica (INAF) and the Istituto Nazionale di Fisica Nucleare (INFN) in Italy. The authors acknowledge support from the INAF grant "Spin and Geometry in accreting X-ray binaries: The first multi frequency spectro-polarimetric campaign"

\clearpage


\clearpage



\begin{table}
 \caption{Polarization degree and angle calculated with {\it ixpeobssim} \cite{18} of the Z-sources observed by \IXPE during the first 1.5 years of observations}
 \label{Z}
\bigskip
\begin{tabular}{|c|c|c|c|}
 \hline
Source Name &{\it PD} (HB) &{\it PD} (NB/FB) & {\it PA} \\
\hline
   --   & \% & \% & $\deg$ ($\deg$) \\
\hline
Cyg X$-$2 & - & 1.85$\pm$0.87 & 140$\pm$12  \\
GX 5$-$1 & 4.3$\pm$0.3 & 2.0$\pm$0.3 & -9.7$\pm$2.0 (-9.2$\pm$4.0) \\
XTE~J1701$-$462 & $(4.6\pm 0.4)\%$ & $<$ 1.5 & -37$\pm$ 2 (-57$\pm$16) \\
\hline
\end{tabular}
\end{table}

\begin{table}
 \caption{Comparison between the spectral parameters of GX 9+9 and GS 1826$-$238}
\bigskip
\begin{tabular}{|l|c|c|}
 \hline
Spectral parameters & GS 1826$-$238 & GX9+9 \\
\hline
$T_{in}$ (KeV) & 0.94$\pm$0.1 & 1.00$\pm$0.04 \\
\hline
$T_{0}$ (KeV) & 1.3$\pm$0.2 & 1.59$\pm$0.07  \\
\hline
$T_{e}$ (KeV) & 2.7$^{+3}_{-0.2}$& 3.4$\pm$0.3 \\
\hline
$\tau$ (-) & 10.8$^{+3.5}_{-6.8}$ & 5.1 $\pm$0.4 \\
\hline
flux$_{2-8 keV}$ (ergs$\times$cm$^{-2}$$\times$s$^{-1}$) & 4.4$\times$10$^{-9}$ & 4.1$\times$10$^{-9}$ \\
\hline
Source Inclination ($\deg$) & 57--61 & 40--60 \\
\hline
Distance (Kpc) & 4--9 & 5--10 \\
\hline
2-8 keV {\it PD}[{\it PA}] (\%[$\deg$]) & $<$1.3[NaN]\footnote{the {\it PA} is not defined if {\it PD} is not constrained} & 1.4$\pm$0.3 [68$\pm$6] \\
\hline
\end{tabular}
\label{gx_gs_tab}
\end{table}

\clearpage


\clearpage



\end{document}